\documentclass[%
 reprint,
superscriptaddress,
 amsmath,amssymb,
 aps,
pra,
]{revtex4-2}

\usepackage{graphicx}
\usepackage{dcolumn}
\usepackage{bm}
\usepackage{physics}
\usepackage{float}
\usepackage{multirow}
\usepackage{gensymb}
\usepackage{svg}
\usepackage{soul}
\usepackage{hyperref}
\usepackage[mathlines]{lineno}
\usepackage{comment}

\emergencystretch=\maxdimen
\begin{document}

\preprint{APS/123-QED}

\title{Violation of local realism with spatially multimode parametric down-conversion pumped by spatially incoherent light}

\author{Cheng Li}
\email{cli221@uottawa.ca}
\affiliation{Department of Physics, University of Ottawa, Ottawa, Ontario, Canada K1N 6N5}

\author{Jeremy Upham}
\affiliation{Department of Physics, University of Ottawa, Ottawa, Ontario, Canada K1N 6N5}

\author{Boris Braverman}
\affiliation{Department of Physics, University of Toronto, Toronto, Ontario, Canada M5S 1A7}

\author{Robert W. Boyd}
\email{rboyd@uottawa.ca}
\affiliation{Department of Physics, University of Ottawa, Ottawa, Ontario, Canada K1N 6N5}
\affiliation{Institute of Optics, University of Rochester, Rochester, New York 14627, USA}


\begin{abstract}
 We experimentally demonstrate a violation of local realism with highly spatially multimode polarization-entangled two-photon states produced by spontaneous parametric down-conversion (SPDC) pumped by a spatially incoherent light source–a light-emitting diode (LED). While existing studies have observed such a violation only by post-selecting the LED-pumped SPDC photons into a single spatial detection mode, we achieve a Clauser-Horne-Shimony-Holt inequality violation of $S = 2.532 \pm 0.069> 2$ using a spatially multimode detection setup that collects nearly 4,080 SPDC spatial modes. These results indicate that coherent pump sources, such as lasers, are not required for SPDC-based entanglement generation. Our work could enable novel and practical sources of entangled photons for quantum technologies such as device-independent quantum key distribution and quantum-enhanced sensing.
\end{abstract}

\maketitle



Entangled quantum systems can violate local realism and exhibit correlations that local hidden variable theories cannot fully explain \cite{Einstein1935PhysRev, Bell1964, Clauser1969PRL}. This feature is arguably a key distinction between classical and quantum systems. In many practical quantum applications, the violation of local realism benchmarks the performance of quantum systems over their classical counterparts. For instance, observing nonlocal correlations between quantum systems certifies the security of a quantum communication channel \cite{Bennett1984, Ekert1991PRL, Mayers1998}. Recent studies have also shown that violation of local realism is a valuable resource for quantum-enhanced imaging and metrology \cite{Moreau2019sciadv, Niezgoda2021PRL, Yadin2021ncomm}. Entangled photons produced from spontaneous parametric down-conversion (SPDC) \cite{Burnham1970PRL, Hong1985PRA, Boyd2020NLO} are excellent candidates for demonstrating such violation \cite{Shih1988PRL, Ou1988PRL, Kiess1993PRL} and enabling practical quantum applications \cite{Jennewin2000PRL, Ursin2007NatPhysics, Black2019PRL, Yin2020Nature, defienne2021natphys}. Coherent light beams, such as those emitted from lasers, have been almost universally employed to pump SPDC and are generally presumed to be indispensable for SPDC-based entanglement generation. Specifically, it has been shown that without postselection, the pump's coherence in a given degree of freedom sets an upper bound on the attainable two-photon entanglement in the \textit{same} degree of freedom \cite{Jha2010PRA, Giese2018PhysicaScripta, Monken1998PRA, Hugo2019PRA, Zhang2019OptExpress, Burlakov2001PRA, Jha2008PRA, Kulkarni2017JOSAB, Kulkarni2016PRA, Meher2020JOSAB}. Recent studies have found that this requirement on the pump's coherence in a given degree of freedom does not necessarily apply to two-photon entanglement in a \textit{different} degree of freedom \cite{Hutter2020PRL}, thus making incoherent light sources such as light-emitting diodes (LEDs) and sunlight possible alternatives to a laser pump. For instance, SPDC pumped by an LED, which is a spatiotemporally incoherent light source, has been shown to produce polarization-entangled photons if the LED light is polarization-filtered \cite{Li2023PRA, Zhang2023PRApplied}. 

However, unlike polarization-entangled photons produced from lasers, those produced from LEDs have not previously been shown to lead to a violation of local realism unless the down-converted photons were first post-selected to reside in a narrow spatial bandwidth using single-mode fibers (SMFs)  \cite{Li2023PRA, Zhang2023PRApplied}. Conducting single-spatial-mode postselection on down-converted photons and detecting them in coincidence effectively projects the pump light into a single spatial mode, thereby defeating the purpose of using a spatially incoherent pump, which contains multiple spatial modes. On the other hand, rejecting photons in higher-order spatial modes effectively reduces detection efficiency and creates a security loophole in quantum communication systems that utilize such devices. We argue that the spatial coherence of the pump beam does not fundamentally limit the attainable polarization entanglement, and the reduced entanglement is a consequence of technical limitations, such as uncompensated walk-offs within the nonlinear crystal that couple the spatial and polarization degrees of freedom \cite{Altepeter2005OptExpress, Rangarajan2009OptExpress, Kuklewicz2004PRA}. In particular, the single-pass, double-crystal setup \cite{Kwiat1999PRA} employed in Ref.~\cite{Li2023PRA} introduces a transverse-momentum-dependent phase between orthogonal polarization components of the pump beam \cite{Li2025arxiv}. Since spatially incoherent light typically has a larger angular bandwidth than spatially coherent light, this phase variation reduces the effective degree of polarization of the pump beam, which limits the maximally attainable two-photon polarization entanglement \cite{Kulkarni2016PRA}. On the other hand, the single-pass, single-crystal setup adapted by Zhang et al. \cite{Kuklewicz2004PRA, Zhang2023PRApplied} introduces a transverse-momentum-dependent phase between orthogonal polarization components of the down-converted photons, which increases their path distinguishability and results in a lower entanglement unless one performs single-spatial-mode postselection. 

Here, we report a violation of local realism with a highly spatially multimode two-photon polarization-entangled state produced from SPDC pumped by the spatially incoherent light beam emitted from an LED. We place the nonlinear crystal inside a polarization Sagnac interferometer (PSI) \cite{Kim2006PRA}. The high degree of symmetry between the counter-propagating directions in the PSI setup can significantly mitigate for temporal and spatial walk-offs that the pump and down-converted photons experience within the crystal. We avoid post-selection into a single spatial mode by coupling the down-converted photons into multi-mode fibers (MMFs) that can collect approximately 4,080 spatial modes of the down-converted photons produced from LED-pumped SPDC. We detect a two-photon state that displays a Clauser-Horne-Shimony-Holt (CHSH) inequality violation of $S = 2.532\pm0.069 > 2$ \cite{Clauser1969PRL}. We quantify the produced two-photon polarization entanglement by conducting quantum state tomography \cite{James2001PRA}, from which we find a concurrence of $C = 0.834\pm0.038$ \cite{Wootters1998PRL}. These results present direct evidence that the coherence of the pump beam in a given degree of freedom does not fundamentally limit the attainable entanglement in a different degree of freedom. In other words, strong entanglement can be produced from incoherent light sources with technical improvements on the setup design. This opens a new avenue for developing novel quantum light sources for practical applications. As a comparison, the same setup pumped by a spatially coherent light beam emitted from a laser yields a CHSH-inequality violation of $S = 2.695\pm0.006$ and a concurrence of $C = 0.952\pm0.002$. The remaining differences between the CHSH-inequality violation and polarization entanglement produced from LED- and laser-pumped SPDC are likely due to wavefront distortions introduced by imperfect optical components. 

\begin{figure}
\includegraphics[width=0.48\textwidth,height=0.48\textheight,keepaspectratio]{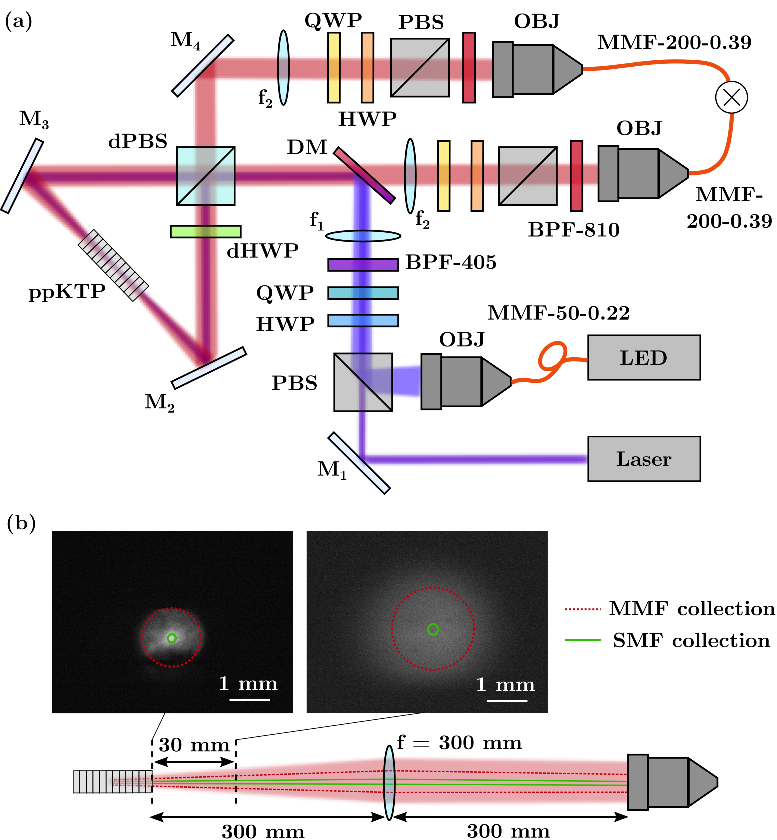}
\caption{\label{fig:1} (a) Schematic of the experimental setup. The SPDC processes for generating polarization-entangled photons occur in the ppKTP crystal, which is placed inside the PSI. M$_{1-4}$: silver mirrors; PBS: polarizing beam splitter; HWP: half-wave plate; dPBS and dHWP: dual-wavelength PBS and HWP; DM: dichroic mirror; ppKTP: periodically-poled potassium titanyl phosphate; QWP: quarter-wave plate; BPF-810(405): band-pass filter centered at 810(405)~nm with a bandwidth of 10(1.5)~nm; MMF-200-0.39(50-0.22): multi-mode fibre with a core diameter of 200(50)~$\mu$m and an NA of 0.39(0.22). OBJ: microscope objective with 20$\times$ magnification and 0.4 NA. (b) Schematic of the procedure for measuring the number of generated and detected spatial modes (not to scale). Inset images show the multimode-laser-pumped SPDC field intensity at z = 0 mm and 30 mm from the PPKTP crystal output face, compared to the effective collection aperture using MMFs (dashed red circles). The solid green circles show the collection aperture of SMFs for comparison.}
\vspace{-1em}
\end{figure}

Fig.~\ref{fig:1} depicts the experimental setup. An LED (Thorlabs M405L3) emits a spatially incoherent pump field with a center wavelength of 405~nm. This light field is first butt-coupled into an MMF with a core diameter of 50~$\mu$m and numerical aperture (NA) of 0.22 (MMF-50-0.22) and subsequently collimated into a beam using a microscope objective with $20\times$ magnification and 0.4 NA. We select the MMF parameters to match the size of the LED pump beam and the transverse dimensions of the nonlinear medium in the same plane, thereby maximizing the interaction region between the pump light and the nonlinear medium. For benchmarking the performance of the setup and comparing the results from pump sources with different coherence properties, a spatially coherent pump beam from an SMF-coupled laser with a center wavelength of 405 nm and a bandwidth of 0.01 nm (Integrated Optics 0405L-23A-NI-AT-NF) is co-aligned with the LED beam into the setup. Using a combination of a polarizing beam splitter (PBS), a half-wave plate (HWP) and a quarter-wave plate (QWP), we can make both pump beams perfectly polarized with arbitrary polarizations, which allows us to maximize the attainable polarization entanglement and generate arbitrary two-photon polarization states residing in the $\{\ket{HV},\ket{VH}\}$ subspace. To characterize LED- and laser-pumped SPDC, we can turn on only the LED or laser and adjust their polarizations to be the same. The pump beam is spectrally filtered using a band-pass filter with a center wavelength of 405~nm and a bandwidth of 1.5~nm (BPF-405) and focused into the nonlinear medium using a lens with a focal length of $f_1 = 200$~mm. After the spectral filtering, the LED beam retains a power of $\sim1.2$~$\mu$W while the laser beam has a power of $\sim1.5$~mW. At the center of the nonlinear medium, the LED pump beam has a diameter of $\sim1$~mm while the laser pump beam has a diameter of $\sim100\ \mu$m. The estimated divergence half-angles for the LED and the laser pump beam are 11 mrad and 2.6 mrad, respectively. 

The nonlinear medium is a 10-mm-long, 2-mm-wide, 1-mm-thick periodically-poled potassium titanyl phosphate (ppKTP) crystal with a grating period of 10~$\mu$m. It is quasi-phase-matched for Type-II collinear frequency-degenerate SPDC from 405 nm to 810 nm at $50\celsius$. We place this crystal inside a PSI comprised of a dual-wavelength PBS (dPBS, Newlight PBS0025-405/810), a dual-wavelength HWP (dHWP, Newlight WPD03-H810-H405), and two silver mirrors. If pump beams propagating in clockwise and counterclockwise directions induce two SPDC processes that are indistinguishable by the detectors, the setup produces a polarization-entangled two-photon state 
\begin{equation}\label{eqn1}
    |\Psi\rangle = \frac{1}{\sqrt{2}}(|HV\rangle+e^{i\phi}|VH\rangle),
\end{equation}
where $\phi$ is the two-photon polarization phase jointly determined by the initial polarization of the pump beam, the phase acquired passing through the dHWP, and the wavefront distortion of the dPBS \cite{Kim2006PRA}. 

The SPDC field is emitted into two distinct paths and collimated by a lens with a focal length of $f_2 = 300$~mm in each. We denote the down-converted photons emitted into the upper path in Fig.~\ref{fig:1} as \textit{signal} while those emitted into the lower path as \textit{idler}. Their joint polarization state is analyzed using a set of QWP, HWP, and PBS in each path. Note that we have chosen all waveplates in the setup to be zero-order (except for the dHWP inside the PSI) and have carefully aligned their optical axes before taking measurements. We use microscope objectives (OBJs) to couple down-converted photons into MMFs with core diameters of 200 $\mu$m and NAs of 0.39 (MMF-200-0.39), which support more than 45,000 spatial modes, and we detect them using avalanche photodiodes (PerkinElmer SPCM-QRH-14-FC). 

Since our photon-collection setup has finite optical apertures, we use the scheme as depicted in Fig.~\ref{fig:1}b to estimate the number of spatial modes collected into the MMFs. To simulate the propagation parameters of LED-pumped SPDC, we couple the laser into the same MMF used for coupling the LED (MMF-50-0.22) and use the output beam from this MMF to pump the crystal. We acquire near-field images of the SPDC at two planes along its propagation path (indicated by vertical dashed lines), one at the output surface of the crystal and the other at 30 mm from the crystal plane. The acquired images are displayed as insets to Fig.~\ref{fig:1}b. By fitting a Gaussian curve to the measured beam profiles, we estimate that LED-pumped SPDC has a diameter of $D_\text{SPDC}^\text{crystal}=1.66$~mm and divergence half-angle of $\theta_\text{SPDC}^\text{crystal}=45$~mrad at the crystal plane. After propagating through the lens $f_2 = 300$~mm, the SPDC profile has an estimated diameter of $D_\text{SPDC}^\text{OBJ}=27.03$~mm and divergence half-angle of $\theta_\text{SPDC}^\text{OBJ}=2.75$~mrad at the OBJ input plane. We estimate the mode field diameter accepted by MMF-200-0.39 at the input plane of OBJ from the numerical aperture of the MMF, which is 0.39, and the effective focal length ($f_e$) of the OBJ we used, which is 10~mm, to be $D_\text{MMF}^\text{OBJ} = 2f_e\tan(\arcsin(\text{NA}))=8.47$~mm, and the corresponding divergence half-angle is $\theta_\text{MMF}^\text{OBJ}=10$~mrad. We note that $D_\text{MMF}^\text{OBJ}$ is smaller than the diameter of the input aperture of the OBJ, which is 10 mm. Since the MMFs accept a larger divergence half-angle than the SPDC field has at the OBJ input plane, we estimate the number of spatial modes of SPDC light collected by each MMF to be $M = (\pi D\theta/\lambda)^2/2=4080$, which indicates that the collected two-photon states are highly spatially multimode. Although the sensor areas of the avalanche diodes (180 $\mu$m) are smaller than the core diameters of the MMFs, we expect to detect all collected spatial modes with approximately the same efficiency because they have undergone mode mixing in the 1-m-long, step-index MMFs. Therefore, our photon-collection setup detects highly spatially multimode SPDC emission, in which the signal-idler correlations truly reflect the influence of the spatially incoherent pump. In contrast, the mode field diameter accepted by SMFs at the input plane of OBJ is approximately 2.62~mm, and the corresponding half-angle is approximately 0.39~mrad, both of which are much smaller than those of the MMFs. For comparison, we depict in Fig.~\ref{fig:1}b the mode field collected by MMFs (dashed red lines) and SMFs (solid green lines).

We extract the photon coincidence rates using a Universal Quantum Devices Logic-16 data-acquisition unit with a coincidence time resolution window $\tau=1$ ns. To test for violation of local realism using the CHSH criterion \(S\leq2\) \cite{Clauser1969PRL}, we measure the polarization correlation between signal and idler photons by fixing the polarization projection in the signal path while recording the coincidence rates as a function of different linear polarization projections of the idler arm. To characterize the resulting two-photon polarization state, we perform quantum state tomography by recording coincidence rates at 16 different polarization projection bases and reconstruct the two-photon density matrix $\rho$, from which we can infer the two-photon polarization phase $\phi$ and quantify the entanglement by calculating the concurrence $C(\rho)$ \cite{Wootters1998PRL}. For LED-pumped SPDC, we set an acquisition time of 5 min for each projective measurement and observe a maximum coincidence rate of $\sim100$ min$^{-1}$ with an accidental coincidence rate of $\sim0.006$ min$^{-1}$; for laser-pumped SPDC, we set the acquisition time for each projective measurement at 10 s and observe a maximum coincidence rate of $\sim11300$ s$^{-1}$ with an accidental coincidence rate of $\sim50$ s$^{-1}$. The accidental coincidence rates are estimated using $2S_sS_i\tau$, where $S_{s(i)}$ is the singles rate in the signal(idler) arm. The average singles rates are $\sim220$ s$^{-1}$ and $\sim155000$ s$^{-1}$ for LED- and laser-pumped SPDC, respectively. We repeat each measurement 10 times to estimate the average value and standard deviation, and subtract accidental coincidence rates from the raw data prior to data analysis. The coincidence rate of LED-pumped SPDC is lower than that of the laser-pumped SPDC for two main reasons: (i) the input power of the LED pump is approximately $10^{-3}$ lower than that of the laser pump; (ii) we estimate the type-II ppKTP crystal to have a narrow phase-matching bandwidth of $\sim$0.3 nm and could only interact with a fraction of the power within the LED pump's 1.5-nm bandwidth to induce SPDC. 

\begin{figure}
\includegraphics[width=0.4\textwidth,height=0.4\textheight,keepaspectratio]{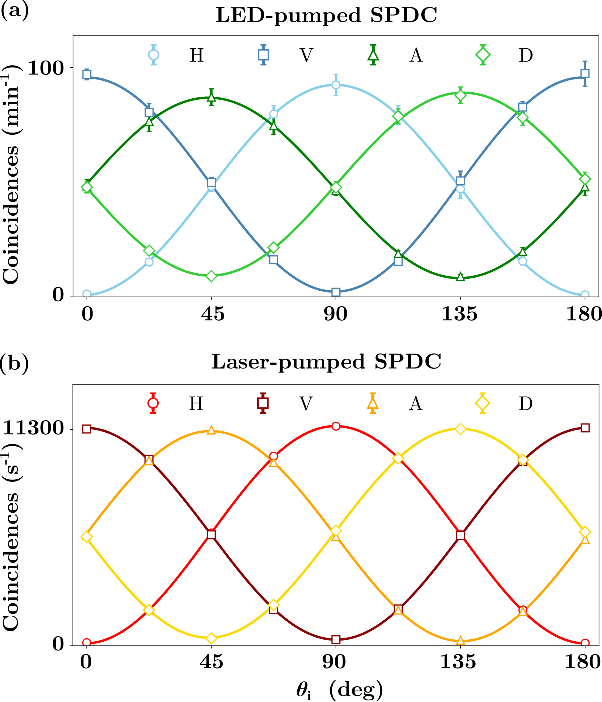}
\caption{\label{fig:2}Polarization correlation fringes of (a) LED-pumped SPDC and (b) laser-pumped SPDC measured by projecting the signal and idler photons into different linear polarization bases represented by angles $\theta_{s(i)}$ with the horizontal polarization. Specifically, we denote the cases of $\theta_{s} = 0^\circ, 45^\circ, 90^\circ, 135^\circ$ as measurements in the horizontal (H), diagonal (D), vertical (V), and anti-diagonal (A) bases, respectively. The markers represent the experimentally measured coincidence rates in H (circles), V (squares), A (triangles), and D (diamonds) bases. The solid lines represent the sinusoidal fitting of the experimental data. The error bars in (b) are indiscernible since they are much smaller than the markers' sizes.}
\vspace{-1.5em}
\end{figure}

Fig.~\ref{fig:2} shows the experimentally measured polarization correlation fringes of LED- and laser-pumped SPDC. We denote $\theta_{s(i)}$ as the angle between the horizontal polarization and the linear polarization bases into which the signal(idler) photons are projected. We then adjust the pump polarization to set $\phi = \pm\pi$ in Eqn.~\ref{eqn1}, thereby targeting $\ket{\psi} = (\ket{HV}-\ket{VH})/\sqrt{2}$, and measure the coincidence rates as a function of $\theta_{i}$ with the signal photons projected into four different bases. Specifically, we choose to project signal photons into the horizontal (H), vertical (V), anti-diagonal (A), and diagonal (D) bases by setting $\theta_{s} = 0^\circ, 90^\circ, 135^\circ\ \text{and}\ 45^\circ$, respectively. The fringe visibility is defined as $(N_\text{max}-N_\text{min})/(N_\text{max}+N_\text{min})$, where $N_\text{max(min)}$ represents the maximal(minimal) coincidence rate measured in each basis. As a result, the LED-pumped SPDC exhibits fringe visibilities of $ 98.07\pm0.70\%$, $ 95.85\pm0.78\%$, $ 81.45\pm1.88\%$, and $81.31\pm2.89\%$ in the H, V, A, and D bases, respectively. The high fringe visibility in the two mutually unbiased bases implies high indistinguishability between the orthogonal polarization components, indicating strong nonlocality even with nearly all spatial modes detected.

Following a similar method as described in \cite{Kwiat1999PRA, Kim2006PRA}, we calculate the CHSH parameter $S$ from the coincidence rates via:
\begin{equation}\label{eqn2}
    S = |E(\theta_s, \theta_i)-E(\theta_s, \theta_i')+E(\theta_s', \theta_i)+E(\theta_s', \theta_i')|,
\end{equation}
and $E(\theta_s, \theta_i)$ is defined as
\begin{equation}\label{eqn3}
\begin{split}
    E&(\theta_s, \theta_i)=\\
    &\frac{N(\theta_s, \theta_i)+N(\theta_s^\perp, \theta_i^\perp)-N(\theta_s, \theta_i^\perp)-N(\theta_s^\perp, \theta_i)}{N(\theta_s, \theta_i)+N(\theta_s^\perp, \theta_i^\perp)+N(\theta_s, \theta_i^\perp)+N(\theta_s^\perp, \theta_i)},
\end{split}
\end{equation}
\begin{equation}
     \theta^\perp =  \theta + 90^\circ,
\end{equation}
where $\theta_{s(i)}^\perp$ stands for an angle perpendicular to $\theta_{s(i)}$. Here, we choose $\theta_{s} = 0^\circ, \theta_{s}' = 45^\circ, \theta_{i} = 67.5^\circ, \theta_{i}' = 22.5^\circ$. Consequently, the LED-pumped SPDC yields $S = 2.532\pm0.069$, which exceeds the classical limit of $S = 2$ by more than 7 standard deviations. To the best of our knowledge, this is the first time that a violation of local realism has been demonstrated with highly spatially multimode two-photon states produced from SPDC pumped by an incoherent light source. More importantly, the CHSH-inequality violation and two-photon polarization entanglement (as will be quantified later by concurrence) measured in this work are significantly higher than those reported in previous works with either MMF or SMF collection of the incoherent-light-pumped SPDC. Table~\ref{tab:table1} summarizes and compares our results with those reported in earlier literature. 

\begin{table*}[!htbp]
\caption{\label{tab:table1}%
Summary and comparison of results in the existing literature on two-photon polarization entanglement produced from SPDC pumped by spatially incoherent light
}
\begin{ruledtabular}
\begin{tabular}{cccc}
\textrm{Reference}&
\textrm{Concurrence}&
\textrm{CHSH-inequality violation}&
\textrm{Photon collection scheme}\\
\colrule
Li et al. (2023)\cite{Li2023PRA} & $0.531\pm0.038$ & N/A & MMF (200 $\mu$m core diameter, 0.39 NA) \\
Zhang et al. (2023)\cite{Zhang2023PRApplied} & 0.63 & $S = 2.33\pm0.097 > 2$ & SMF\\
 & 0.54 & N/A & MMF (50 $\mu$m core diameter)\\
 & 0.35 & N/A & MMF (125 $\mu$m core diameter)\\
This work & $0.834\pm0.038$ & $S = 2.532\pm0.069 > 2$ & MMF (200 $\mu$m core diameter, 0.39 NA)\\
\end{tabular}
\end{ruledtabular}
\end{table*}

In contrast, the laser-pumped SPDC exhibits fringe visibilities of $ 97.69\pm0.07\%$, $ 94.62\pm0.10\%$, $ 95.08\pm0.07\%$, and $93.40\pm0.09\%$ in the H, V, A, and D bases, respectively. The corresponding S-parameter is calculated as $S=2.695\pm0.006$, displaying a violation of local realism by more than 115 standard deviations. While LED- and laser-pumped SPDC exhibit comparably high fringe visibility in the H-V bases, the laser-pumped SPDC results in a notably higher fringe visibility in the A-D bases and a stronger violation of local realism compared to LED-pumped SPDC. A reduced fringe visibility in the mutually unbiased basis can be attributed to two potential reasons: (i) the produced state has an additional phase between the $|HV\rangle$ and $|VH\rangle$ components compared to the target state $|HV\rangle-|VH\rangle$ so that the fringes measured in the A-D bases have reduced visibilities; (ii) the produced state has reduced entanglement, which manifests as a reduced indistinguishablity in the mutually unbiased bases. To fully characterize the resulting two-photon states produced from different pump sources, we conduct quantum state tomography to reconstruct their density matrices $\rho$ and calculate $C(\rho)$ to quantify the entanglement.  

\begin{figure}[b]
\includegraphics[width=0.4\textwidth,height=0.4\textheight,keepaspectratio]{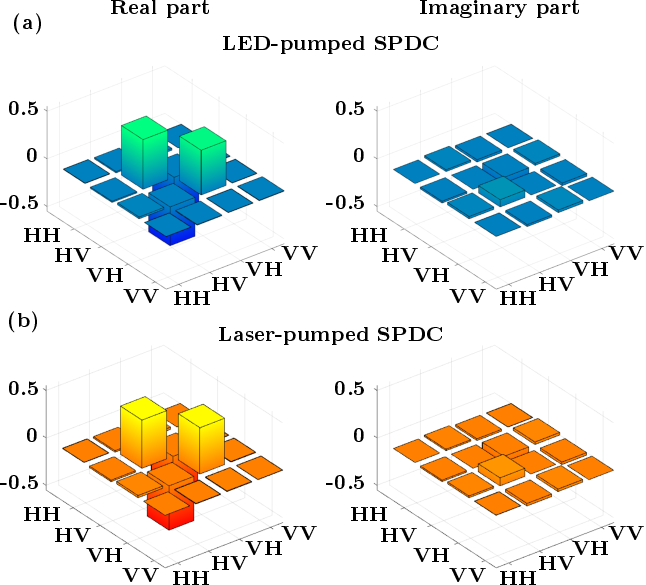}
\caption{\label{fig:3}Real and imaginary parts of the density matrices of the two-photon states produced with (a) LED-pumped SPDC and (b) laser-pumped SPDC. All elements of the density matrices have uncertainties less than 0.03 for LED-pumped SPDC and less than 0.003 for laser-pumped SPDC.}
\vspace{-1em}
\end{figure}

In Fig.~\ref{fig:3}(a) and (b), we depict the real and imaginary parts of the two-photon density matrices for LED- and laser-pumped SPDC, respectively. By calculating the phase of the $\ket{VH}\bra{HV}$ element, we infer that the LED-pumped SPDC produces a state with $\phi = (-0.941\pm0.024)\pi$ and the laser-pumped SPDC produces a state with $\phi =(-0.943\pm0.001)\pi$. The fidelity of the experimentally measured state to the target state is $89.88\pm0.51\%$ for LED-pumped SPDC and $96.36\pm0.04\%$ for laser-pumped SPDC, respectively. These results indicate that we have chosen a near-optimal set of polarization projection bases. The two-photon state produced from LED-pumped SPDC has a concurrence of $C = 0.834\pm0.038$, whereas the one produced from laser-pumped SPDC has a concurrence of $C = 0.952\pm0.002$. The lower value of $C$ observed for the LED-pumped SPDC means the produced state has a reduced entanglement compared to laser-pumped SPDC, while both produce non-maximal entanglement \cite{Wootters1998PRL}. Therefore, the lower violation of local realism of LED-pumped SPDC is mainly due to the reduced entanglement in the generated state. The concurrence results are consistent with the differences in fringe visibilities measured in the A-D basis since the entanglement is sensitive to the indistinguishability between the SPDC processes induced by pump light traversing the PSI in the clockwise and counterclockwise directions. 

The non-maximal entanglement produced from both types of pump sources can be attributed to the wavefront distortions within the PSI. Specifically, the uneven distances between the two cemented prisms comprising the dPBS cause the H- and V-components of both the pump and the down-converted photons to acquire different relative phases across the transverse plane. This effect is analogous to a waveplate with a spatially non-uniform phase retardation, which alters the polarization state of the light in a position-dependent manner. The influence of wavefront distortion experienced by the down-converted photons is discussed in Ref.~\cite{Kim2006PRA} in the context of laser-pumped SPDC. Collecting the down-converted photons using MMFs effectively mixes highly entangled states (Eqn.~\ref{eqn1}) with different $\phi$, thereby reducing the overall polarization entanglement. 

\begin{figure}
\includegraphics[width=0.48\textwidth,keepaspectratio]{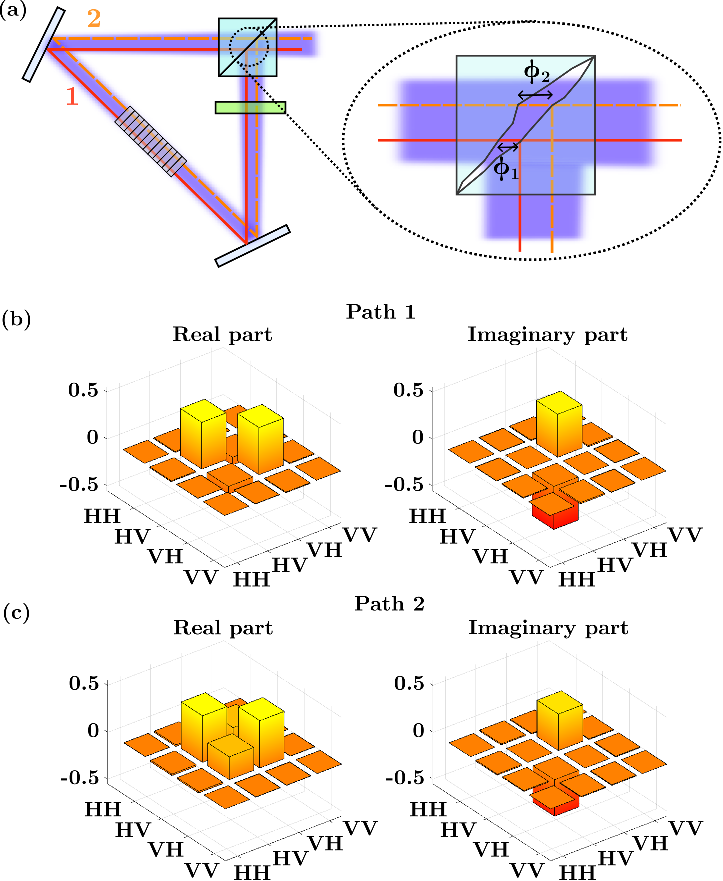}
\vspace{-1.5em}
\caption{\label{fig:4}(a) Schematic diagram of probing the effects of wavefront distortion on the produced two-photon state. The laser pump beam enters the PSI at two alternative paths with a horizontal shift of $\sim1$~mm. The inset depicts the effect of wavefront distortion, which introduces different polarization phase retardation at different transverse positions of the pump beam. The uneven distances between the two cemented prisms are exaggerated for illustration purposes. The reconstructed density matrices of two-photon states produced in (b) path 1 and (c) path 2.}
\vspace{-1em}
\end{figure}

In the case of laser-pumped SPDC, the influence of wavefront distortion on the pump beam is negligible since the laser beam's spot size is sufficiently small that the phase change appears uniform across its transverse profile. However, this effect becomes more prominent for LED-pumped SPDC. We depict the effect of wavefront distortion on the LED pump beam in Fig.~\ref{fig:4}(a). Since the light beam from an LED has a larger spot size than that of a laser, the pump beam picks up spatially varying polarization phase across its transverse profile. For instance, as shown in the inset to Fig.~\ref{fig:4}(a), pump photons entering the PSI through path 1 (solid line) acquire a relative phase of $\phi_1$ between their H- and V-components, while those entering through path 2 (dashed line) acquire $\phi_2$, with $\phi_1\neq\phi_2$ in general. As a result, the down-converted photons generated from different transverse components of the LED pump already carry different two-photon polarization phases $\phi$ before undergoing further distortions at the dPBS. Therefore, the LED-pumped SPDC is a mixture of two-photon states with different $\phi$, which are generated from pump photons of different transverse positions. As a result, the detected states display reduced concurrence when averaging over all possible transverse positions. To characterize this effect, we adjust the alignment of the laser pump beam so that it can traverse the PSI in two alternative paths separated by $\sim 1$ mm in the horizontal direction, as depicted in Fig.~\ref{fig:4}(a). Consequently, the influence of wavefront distortion on the pump beam can be observed by comparing the resulting two-photon states produced from laser-pumped SPDC induced in different paths. In Fig.~\ref{fig:4}(b) and (c), we show the density matrices of the two-photon density matrices produced in paths 1 and 2, respectively. The two-photon state produced by the pump beam traversing path 1 yields $C=0.933\pm0.001$ and $\phi= (0.5558\pm0.0002)\pi$, while that produced in path 2 exhibits $C=0.916\pm0.001$ and $\phi=(0.3220\pm0.0001)\pi$. In both cases, the two-photon state displays strong ($C>0.9$) entanglement, indicating that the SPDC processes induced in the clockwise and counter-clockwise directions are sufficiently indistinguishable along either path. The observed non-maximal entanglement is primarily due to the wavefront distortion experienced by the down-converted photons, which is similar to the results depicted in Fig.~\ref{fig:3}(b) and those reported in \cite{Kim2006PRA}. The two output states display a distinct two-photon polarization phase $\phi$ due to the pump beam entering the PSI at different transverse positions on the dPBS. Using a pump beam with a larger spot size effectively mixes the non-maximal two-photon states with different $\phi$ produced in different paths from pump beams with smaller spot sizes, which further reduces the resulting entanglement. For instance, an equal mixture of the two states depicted in Fig.~\ref{fig:4}(b) and (c) has a lower concurrence of $C=0.8558\pm0.0003$, which is comparable to that of the LED-pumped SPDC. These measurements suggest that the spatial coherence of the pump does not fundamentally limit the polarization entanglement produced from SPDC, and technical improvements can further enhance the produced entanglement.

In summary, we have demonstrated a violation of local realism with a highly spatially multimode two-photon state produced from SPDC pumped by a spatially incoherent light source—an LED. By collecting the down-converted photons with MMFs, we achieve a violation of the CHSH inequality $S = 2.532\pm0.069 > 2$ by more than 7 standard deviations using a two-photon state containing nearly 4080 spatial modes. Tomography analysis shows that the two-photon state produced from LED-pumped SPDC has a concurrence of $C = 0.834\pm0.038$. These results represent the strongest violation of local realism and entanglement reported to date for incoherent-light-pumped SPDC. In contrast, SPDC pumped by a spatially coherent laser beam in the same setup results in $S = 2.695\pm0.006$ and $C = 0.952\pm0.002$. The reduced local realism violation and entanglement observed for LED-pumped SPDC are likely results of the wavefront distortion introduced by the imperfect optics parts within the PSI. We demonstrate this by observing that near-maximal entangled states produced from laser pump beams traversing different paths in the PSI are distinguishable by their two-photon polarization phase. 

Our results inform future studies employing incoherent-light-pumped SPDC for practical entangled photon sources. For instance, one can compensate the wavefront distortions experienced by the pump and down-converted photons with additional phase masks or adaptive optics devices \cite{Hugo2018PRL, Cameron2024Science} to produce maximal entanglement from spatially incoherent light. The viability of producing high entanglement regardless of the pump's coherence opens up new opportunities for entanglement-based quantum information technologies. For instance, since incoherent light sources such as LEDs and sunlight are more ubiquitous and less power-consuming than most coherent light sources such as lasers, incoherent-light-pumped SPDC could be employed to build robust entangled photon sources without active electronics (other than temperature control of the nonlinear cystals), which extends the accessibility of quantum technologies in resource-restricted environments such as the Arctic region and satellites in space \cite{Yin2020Nature}. Furthermore, two-photon polarization-entangled states produced from SPDC pumped by incoherent light may be more suitable for free-space quantum key distribution due to their robustness against atmospheric turbulence \cite{Bhattacharjee2020OptLett, Qiu2012APB, Phehlukwayo2020PRA}. In the broader context of entangled photon generation using nonlinear optical methods, a recent study demonstrated that four-wave mixing driven by amplified spontaneous emission can produce highly entangled two-photon states with enhanced pair generation rate \cite{song2025arxiv}, underscoring the potential of incoherent-light-driven quantum information technologies with nonlinear optical platforms.

C. L. acknowledges useful discussions with Yingwen Zhang, Alessio D'Errico, Yang Xu, and Jiapeng Zhao. The portion of the work performed at the University of Ottawa was supported by the Canada Research Chairs program under Award 950-231657, the Natural Sciences and Engineering Research Council of Canada under Alliance Consortia Quantum Grant ALLRP 578468 - 22, Discovery Grant RGPIN/2017-06880, and the Canada First Research Excellence Fund Award 072623. In addition, R.W.B. acknowledges support through U.S. National Science Foundation Award No. 2138174 and U.S. Department of Energy Award No. FWP 76295.

\nocite{*}

\bibliography{ms}
\end{document}